\documentclass[aps,prl,twocolumn,nopacs,amssymb]{revtex4}

\usepackage[dvips]{graphicx}
\usepackage{dcolumn}
\usepackage{bm}
\def\be{\begin{equation}}
\def\en{\end{equation}}

\def\p{\partial} 
\newcommand{\av}[1]{\left\langle{#1}\right\rangle}

\def\gs{\gtrsim}
\def\ls{\lesssim}
\newcommand{\bi}[1]{\mbox{\boldmath$#1$}}
\def\p{\partial}
\def\bea{\begin{eqnarray}}
\def\ena{\end{eqnarray}}

\begin{document}


\title{Acoustic resonance   in  periodically sheared  glass}



\author{Takeshi  Kawasaki$^1$ and Akira Onuki$^2$}
\affiliation{
$^1$Department of Physics, Nagoya University, Nagoya 464-8602, Japan\\
$^2$Department of Physics, Kyoto University, Kyoto 606-8502, Japan\\ 
}


\date{\today}

\begin{abstract}
Using molecular dynamics simulation,  we study acoustic  
 resonance  in low-temperature glass by applying a  small  periodic shear 
at a boundary wall. Shear wave resonance   occurs 
as  the  frequency $\omega$ approaches   
$\omega_\ell= \pi c_\perp\ell/L$ ($\ell=1, 2, 3,...)$. 
Here,  $c_\perp$ is the transverse sound speed and $L$ is the 
cell length. At resonance,  large-amplitude  sound waves appear  
after many cycles even for very small applied strains.  
They then induce  plastic events, 
which  are   heterogeneous in space and   intermittent 
on time scales longer  than  the oscillation  period $2\pi/\omega$. 
From these irreversible particle motions, there arises strong  
dissipation suppressing the growth of  sounds. 
After many resonant cycles, we observe a phenomenon of forced aging, 
where the shear modulus (measured after switching off 
the  oscillation) is increased  significantly.
 Sometimes, exceptionally 
large   plastic events  and system-size  sliding motions 
induce   a transition  from  resonant to 
 off-resonant states. At resonance,  translational 
diffusion becomes appreciable  as well as aging due to enhanced 
configurational changes.
\end{abstract} 


  

\maketitle


{\it Introduction.--}
Systems with oscillating degrees of freedom can 
resonate to  an externally applied periodic perturbation 
  as its frequency $\omega$ 
approaches  a resonance  one $\omega_r$   \cite{Landau,Mook,Nature}. 
In fact,  parametric resonance has  been 
observed in various systems   
with  spin waves \cite{Suhl} and 
surface waves \cite{Faraday}. It is   
well  known   that small-amplitude  mechanical 
perturbations can greatly  excite particular 
sound modes in many systems (including musical instruments). Such  
acoustic resonance has been  used to accurately determine 
  the elastic moduli \cite{RUS}, when  the resonance width  
$\Delta\omega$ 
is sufficiently small in the frequency range. 
In  crystals,  dislocation motions 
give rise to damping of  large-amplitude sounds, 
so   $\Delta\omega$ should depend on the defect density 
\cite{damp}.  For  fluids, 
we should include the  transport coefficients  and   the nonlinear 
terms in the hydrodynamic equations to describe 
resonance of  longitudinal sounds. 
In particular, in fluids near  gas-liquid criticality, 
 resonance   saturation is due to  
  the singular    bulk viscosity  \cite{Moldover}. 

In this Letter, we report unique aspects 
of   acoustic resonance in glass at low temperature $T$.  
 Here, we should mention recent papers   on  
 dynamics of   glass   under periodic shear in the low frequency limit 
\cite{Pine,Keim,Priez,Hern,Regev,Sood,Sastry,Berthier,Schall,Haya}.  
  These papers have confirmed  
 that  the particles motions can be    microscopically reversible 
for small strain  amplitude $\gamma_0$  but become 
partially irreversible with increasing $\gamma_0$ at low $T$. 
 In contrast, as  $\omega \to \omega_r $ 
with small $\gamma_0$, the  energy input from   a wall 
 accumulates in the cell even if it is small in one cycle. 
Thus, after many cycles, 
 there appear  regions with relatively large strains, 
where  plastic events  occur  
 heterogeneously  and intermittently  on time scales longer than the period 
$t_p=2\pi/\omega$ \cite{Yamamoto,Anael}. 
Inducing    random particle motions and emission of sounds \cite{Shiba}, 
they give rise to a dissipation mechanism, 
which suppresses  the  growth of sounds 
 and determines $\Delta \omega$.

Between two parallel walls with  distance $L$,  
the reflection time of shear waves is $t_r =  2L/c_\perp$, 
where  $c_\perp$ is the transverse sound speed. 
If one wall is oscillated at a small $\gamma_0$, shear wave 
 resonance occurs for  $t_r = \ell t_p$  
or for $\omega = \pi \ell  c_\perp/L$  ($\ell=1, 2, \cdots)$, 
where the wave nodes are at the walls.  
However, this criterion is only approximate   
because of the following. First,  the sound  modes in glass 
are highly heterogeneous   
 and the continuum theory  holds only at 
 very long wavelengths 
 \cite{Gelin,Sch,Elliott,Monaco,Barrat,Reichman,KawaJCP}. 
Second, the amplified   sound waves at resonance are largely  
deformed from sinusoidal forms, where 
plastic events are proliferated and 
 the linear elasticity does not hold.

In   amplified sounds in glass, 
the  particles should  noticeably jump out of  cages.  
We shall indeed detect enhanced diffusion at resonance.   
Moreover, if the system is at resonance 
 for a long time, there should be  acceleration of 
the aging processes (which are  extremely slow in quiescent states) 
\cite{Nagel,Lacks}.  In fact, we shall find  
a significant increase in  the shear modulus $G$ 
after many resonant cycles. 
This effect may be called {\it resonance hardening}.

 {\it Simulation method.--} 
Our system is  a two-dimensional 
binary mixture  in glassy states. In a $L\times L$  cell, 
the particle  numbers  are 
$N_1=N_2=N/2$ with $N=4000$.
The particle pairs separated by 
$r$   interact  via  potentials, 
\be 
\phi_{\alpha\beta} (r) = 
\epsilon  ({\sigma_{\alpha\beta}}/{r})^{12} -
C_{\alpha\beta} 
\quad (r<r_{\rm c}), 
\en 
where we introduce   $\epsilon$, 
$\sigma_{1}$, and  $\sigma_2=1.4\sigma_1$ with  
  ${\sigma_{\alpha\beta}}= ({\sigma_{\alpha}}+{\sigma_{\beta}})/2$. 
Here, $\phi_{\alpha\beta}=0$ 
for $r\ge r_{\rm c} =4.5 \sigma_{\alpha\beta}$ 
with the constant $C_{\alpha\beta}$ ensuring  the continuity of 
 $\phi_{\alpha\beta}$ at  $r=r_{\rm c}$.  
The   mass ratio is  $m_1/m_2=1.96$. We will measure 
space, time, and temperature   in units of $\sigma_1$, 
$t_0= \sigma_1(m_2/\epsilon)^{1/2}$, and $\epsilon/k_B$, respectively. 
Then, the cell length is   $L=70.2$. 

To  the cell ($0<y<L$), 
we attached  two boundary layers in the regions 
$-L/16<y<0$ and $L<y<17L/16$. Each layer contains 
250 particles  bound  to  pinning points   ${\bi R}_j$ on it 
 by   the spring potential 
$\phi_j = 100\epsilon |{\bi r} -  {\bi R}_j|^2/\sigma_1^2$,  
where  ${\bi R}_j$ were   determined  in a liquid state \cite{Shiba}. 
These boundary particles interact  with  those in the cell 
  via  the potentials in Eq.(1), so  layer motions along the $x$ axis induce 
 shear motions  in the cell. Keeping  the lower layer    at rest, we moved 
 the upper one   along the $x$ axis as    
\be
u_x(L,t)= -d \sin (\omega t)\quad (t>0),   
\en 
where $d$ is a small displacement. 
 In this Letter, the mean applied strain    
$\gamma_0= d/L$  is very small (0.002 
for $d=0.15$).

To  prepare initial glassy  states,
we started with  a liquid  at a high $T$, lowered $T$  
 to  $0.01$  below the 
glass transition, and waited  for a time of  $10^3$, 
where we used  Nos\'e-Hoover thermostats in the three space regions. 
After these steps, we  removed  
 the  thermostat in the cell at $t=0$, keeping those in the boundary layers. 
Using this initial state for each $d$ and 
$\omega$,   we    applied the shear  in Eq.(2); then,  
  the local temperature (the local average of 
the kinetic energy per particle) became 
inhomogeneous due to heating, but it  was  fixed 
 at $0.01$ in the boundary layers.

\begin{figure}
\includegraphics[scale=0.6]{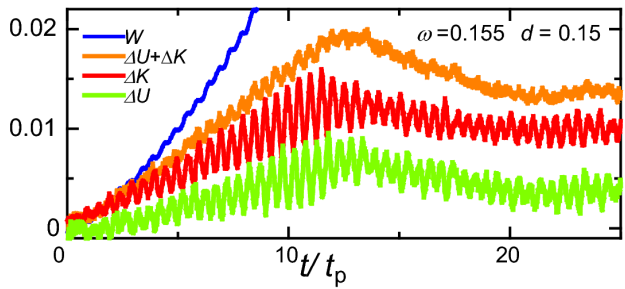}
\caption{(Color online) Evolution after 
application of  periodic shear with  $d=0.15$ 
at  first resonance frequency $\omega=\omega_1 =0.155$. 
Here,  $0<t<25t_p$ with  $t_p=40.5$. 
Plotted are  deviations of 
  kinetic energy $ K$,   potential energy $U$, 
and   sum $K+U$ from their initial values 
divided by  $N\epsilon $.    $W(t)$  (in blue) is  energy input  from 
the upper boundary layer in units of $N\epsilon$.  
 }
\end{figure}

 {\it Resonance.--}  In this Letter,  the  resonant  
frequencies are close to $ \omega_\ell= \pi \ell c_\perp/L$ 
up to $\ell=3$. The latter  are the frequencies 
  of the standing shear  waves. In our 
 initial state, we have  $c_\perp=(G/\rho)^{1/2}= 3.5$,  
where $G=16~\epsilon/\sigma_1^2$ 
is the  shear modulus and  $\rho=1.20~{ m}_2/\sigma_1^2$ is the  mass density. 
In  Supplementary Material (SM) \cite{Supp},  
we  present microscopic  analysis of the  vibrational 
modes  \cite{Gelin,Sch,Elliott,Monaco,Barrat,Reichman}, 
 where the first one with the lowest 
frequency represents  the    shear wave with $\ell=1$ 
and  the quasi-localized ones  have higher frequencies 
 for our system size.  
In SM \cite{Supp}, 
 we also provide   a movie of  
resonant growth.

At the first resonance $\omega =\omega_1= 0.155$  with  $d=0.15$, 
 Fig. 1 displays  growth of  the kinetic energy 
$K(t)$,  the potential energy $ U(t) $, and their  sum $H(t)$ 
of  the particles  in the cell. 
 The deviations 
$\Delta K(t)=K(t)-K(0)$ and $\Delta U(t) =U(t)-U(0)$ 
from the initial values  consist  
of oscillating  parts due to sounds and slowly evolving parts 
due to  heating.  
 The sum of the former is the total acoustic energy with  weaker  
oscillation, which grows   
 up to  $ 0.005N\epsilon$  for $d=0.15$ and 
$ 0.02N\epsilon$  for $d=0.3$.  
The    temperature    in the middle 
 is  higher than  0.01  
by $ 0.025$ for $d=0.15$  and by  $0.04$ for $d=0.3$ 
  for    $t/t_p\gs  20$.  We also plot the   energy input from the 
upper  layer  to the cell, denoted by  
$W(t)$ (see its definition 
in Ref. \cite{input}). It     is initially 
changed into    the acoustic energy 
but  is eventually balanced with the energy transport 
from the cell   to the boundary layers \cite{Shiba}.

We next examine how the resonance
occurs as  $\omega$ is varied. 
We define   the  average displacement length by 
\be 
{\Delta r}(\omega,\tau)= \hspace{-2mm}
\sum_{0\le n -n_0<M} \sum_{i} 
\frac{ |{\bi r}_i(nt_p+\tau) -{\bi r}_i(nt_p)|}{M N},  
\en 
where $0< \tau\le t_p$. We  sum  over 
the particles in the cell 
and over $M$ consecutive cycles ($n_0 \le n<n_0+ M)$.  With  
 $n_0=50$ and $M=200$, Fig. 2 gives    
${\Delta r}(\omega,\tau)$ vs $\omega$ for  
 (a) $\tau=t_p/4$ and (b) $\tau=t_p$.
 The displacements in (a) consist of  reversible (periodic) 
and irreversible (non periodic) ones, while those in (b) are all  irreversible. Amplification  occurs around   $\omega\sim  0.15, 0.33$, and $0.53$, 
which  correspond to $\omega_\ell$ ($\ell=1,2,3$). 
For  $\omega\sim 0.15$, the reversible ones  are dominant 
such that ${\Delta r}$ in (a) 
 is  much larger  than  ${\Delta r}$ in (b).  
For  $\omega \sim 0.33$ and 0.53, 
the resonance width $\Delta\omega$ is large 
with enhanced  irreversibility. 
In  Fig. 3, we display  typical  amplified  
displacements in a quarter period  ($n<t/t_p< n+1/4$), 
where $\omega=0.155$, 0.325,  and 0.525 with    $d=0.3$. 
These  correspond to the first three shear waves, but    
   they are    deformed from sinusoidal forms and their irregularity  is 
more marked for larger $\omega$.

\begin{figure}[t]
\includegraphics[scale=0.5]{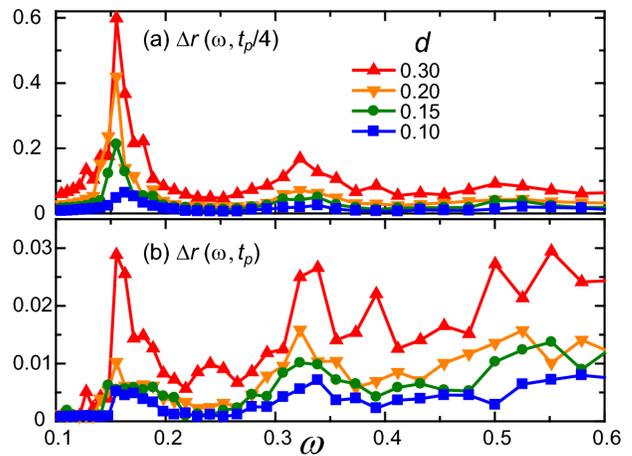}
\caption{(Color online) Average 
displacement length  ${\Delta r}(\omega,\tau)$ in Eq.(3) vs $\omega$  
for (a) $\tau=t_p/4$ and  (b) $\tau=t_p$  with 
$d=0.3,0.25, 0.2$, and 0.1. Average is taken 
over 200 cycles for $t \ge  50t_p$. 
At $\omega \cong \omega_1$ 
 reversibility is conspicuous 
in (a), but irreversibility  increases 
 with increasing $\omega$ in (b).  
 }
\end{figure}

\begin{figure}
\includegraphics[width=0.97\linewidth]{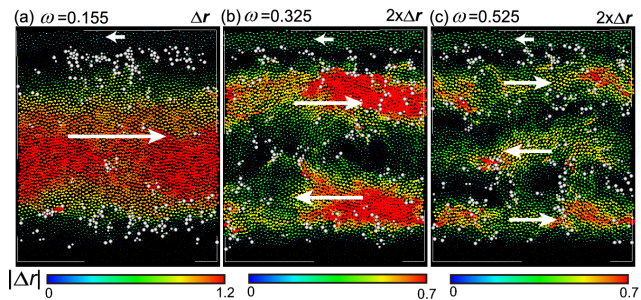}
\caption{(Color online) Displacements 
$\Delta {\bi r}_i={\bi r}_i(nt_p+t_p/4)-{\bi r}_i(nt_p)$ 
in a quarter cycle  at $n \sim 100$  for  $d=0.3$, 
where $\omega$ is (a) $\omega_1 
= 0.155$,  (b) $\omega_2=0.325$,   and (c) $\omega_3=0.525$. 
Colors represent 
$|\Delta {\bi r}_i|$ according to the color bar. 
Particles in white are those  with bonds broken  in 
 $[nt_p, (n+1)t_p]$.    
White arrows are eye guides 
to show the  average of   $\Delta {\bi r}_i$  along the $x$ axis. 
}
\end{figure}

To describe  plastic events, we here 
introduce the bond breakage   \cite{Yamamoto} for each cycle. Namely, 
 particles $i$ and $j$ have broken bonds  if their 
 distance $r_{ij} (t) $ 
is shorter than $ 1.15 \sigma_{\alpha\beta}$ at $t=nt_p$ 
and is longer than 
$1.25 \sigma_{\alpha\beta}$ at $t=(n+1)t_p$.   
In  Fig. 3,   these particles  are marked (in white). Then 
plastic events are  visualized, 
which are    collective and heterogeneous, 
taking  place  more frequently  
in regions with larger velocity gradients. 
 Let  $N_{\rm B}(\omega, n)$  be  
 the number of these particles with broken bonds in the $n$-th cycle.
 Then,   the energy dissipation at resonance  
fluctuates around $\epsilon  N_{\rm B}(\omega, n)$ in each  cycle 
 \cite{comment1}. 
In fact, the averages of the 
 one-cycle  energy input $\Delta W(n)= W(nt_p+t_p)-W(nt_p)$ \cite{input}  
and  $\epsilon  N_{\rm B}(\omega, n)$ 
over $n$  are both about $50\epsilon$ for $\omega=0.155$ and 
$d=0.3$.

 {\it Intermittency and big drop.--} 
In Fig. 4(a),   we show $N_{\rm B}(\omega, n)$ vs $n$ 
 at $\omega=0.155$  in the range   $n= t/t_p\le 1300$.  
 See  its  behavior on  shorter time  scales     
in SM \cite{Supp}. It  evolves intermittently 
for $n\ls 800$  but  largely drops 
 at  $n\sim 800$. This drop  indicates a transition  
from  resonant    to  off-resonant  states, which 
 is similar to the  absorbing transitions  
from   active    to  inactive  states  
\cite{Abs,Pine,Keim,Berthier,Sood}. 
In (b), we plot   the energy deviation 
$\Delta H(t) = \Delta K+\Delta U$  from its  initial value 
at $\omega=0.155$, whose fluctuations greatly increase 
with increasing  $d$.    For $d=0.2$ and $0.3$, 
it  drops to   negative values ($-0.01N\epsilon $ 
and $-0.02 N \epsilon $, respectively).
The curves  of $d=0.3$ in (a) and (b) are obtained from the same run.  
With  the drop,  cooling occurs 
to the boundary temperature 0.01 
on  time scales  of $25-50 t_p$.  
On the other hand, for $d=0.15$, $\Delta H$   decreases 
 to a small positive value ($\sim 10^{-3}$), where   
a  weakly resonant state follows. 
The time of these transitions is random depending on the 
initial state.

\begin{figure}[t]
\includegraphics[scale=0.5]{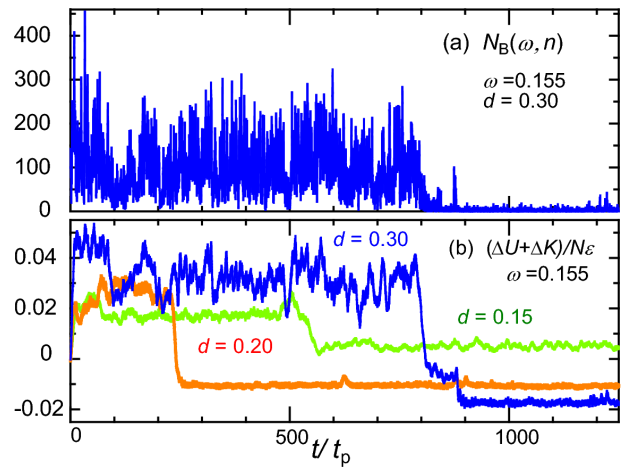}
\caption{(Color online) 
Long time behaviors    at $\omega=0.155$ for $n= t/t_p<1300$.
(a) Particle  number $N_B(\omega, n)$ with broken bonds  vs $n$ 
 for   $d=0.3$, showing intermittency  for $n\ls 800$. 
(b) Normalized deviation of    energy  $\Delta H/N\epsilon = 
(\Delta K+\Delta U)/N\epsilon$.  
  Resonance disappears for  $d=0.3$ at $n \sim 800$ and 
for  $d=0.2$ at $n \sim 350$, while it is weakened for  $d=0.15$ 
at $n\sim 550$.  
 }
\end{figure}

 {\it Forced aging.--}  
We show that the aging is 
accelerated during resonance \cite{Lacks,Nagel}. 
In  Figs. 1 and 4(b), however, heating and  amplified waves 
yield positive energy changes   $\Delta H>0$  
from the initial value. Thus,  we switched off 
the oscillation after many resonant cycles and   
cooled the cell  to 0.01 
in  an equilibration   time of $10^3$.   
If we use   the data of $d=0.3$ in Fig. 4   after $M$  cycles, 
 $\Delta H$ is   $-0.045N\epsilon$ for $M=600$ and is $-0.047N\epsilon$
for $M=1000$ after cooling.  Furthermore, in Fig. 5(a), the shear modulus 
 $G$ from the stress-strain relation 
(in units of $\epsilon\sigma_1^{-2}$) 
 is  20  both  in these two cases after cooling, which is considerably 
larger than the initial value   $ 16$.  
Since   the big drop is at  $t/t_p \sim 800$ for $d=0.3$ in Fig. 4, 
the structure change leading to  this  
hardening  should have occurred   before  the big drop. 
If we again applied  a periodic shear to  these cooled states, 
 resonance occurred  at 
a higher frequency  about  $\pi(G/\rho)^{1/2}/L\cong 0.17$ 
(not shown here). Thus, the resonant states realized in simulation  
are history-dependent. For the run of $d=0.15$  in Fig. 4(b), 
$G$  increased only by  1 from its initial value.

At  high-amplitude resonance, the waves are largely deformed  
on mesoscopic scales in considerably heated 
regions, where $G$ considerably 
depends on $T$  \cite{comment2}.   Thus, the 
  sound propagation  in resonance  
is  very complicated. 
Remarkably, at big drops breaking resonance,  
 we observed  exceptionally large plastic events 
and system-size  sliding motions, as in Fig. 5(b).   
 We conjecture that these  large-scale  motions 
 break the resonance condition.  
In Fig. S4 in SM \cite{Supp}, we will visualize smaller-scale 
sliding  motions  not breaking resonance. 
As a similar finding, 
Fiocco {\it et al.} \cite{Sastry} numerically realized  a 
 thick shear band  at large periodic strains. 
It is worth noting  that long-range elastic 
deformations are produced around local plstic events \cite{Anael}.
We should further study 
these large-scale  motions (in addition to 
plastic events) in sheared glass. 

 {\it Diffusion.--}
The particles can jump  out of cages 
appreciably  at large strains even at  
 very low  $T$ \cite{Berthier,Sastry,Priez}. 
This is consistent with 
 our claim  that the aging 
processes are accelerated at resonance. 
Here, we  examine  the stroboscopic 
 mean square displacement along the $y$ axis in time intervals with width 
$nt_p$ written as 
\be 
M(n) = \av{\sum_{1 \le i\le N}  |{y}_i((n+s)t_p)-{y}_i(s t_p)|^2/N}. 
\en   
where the average is taken over $s$  
($0\le s<100$) at  fixed $n$.  In Fig. 6,  we plot 
$M(n) $   in the range  $n<10^3$.  
 For $d=0.3$, it  grows  as $4D_\perp nt_p$ for $n\gs 20$, 
where the diffusion constant 
is given by $D_\perp= 5.0 \times 10^{-5}~\sigma_1^2/t_0$  
\cite{diffusion}.

For $d=0.15$ in Fig. 6,  
$M(n)$  remains close to its  plateau, so it does  
not give   $D_\perp$. Here, we note that 
the diffusion constant  can  be  obtained from 
short-time analysis of jump motions  in   glass  \cite{KawasakiPRE}. 
To this end,  we 
 pick up the particles 
with large displacement $|{\bi r}_i((s+n)t_p)-{\bi r}_i(st_p)|>\ell_{\rm J}=
0.8$, where $\ell_{\rm J}$ gives 
the first minimum of the Van Hove self-correlation function. 
Their contribution to $M(n)$ in Eq.(4) 
is written as 
\be 
M_{\rm JP}(n)= 
\av{\sum_{i\in {\rm jump}}  |{y}_i((n+s)t_p)-{y}_i(s t_p)|^2/N}. 
\en   
where  we remove  the contribution from    
the thermal  cage motions. Indeed, for $d= 0.15$ in  Fig. 6, we find 
$M_{\rm JP} (n) \cong  4D_\perp nt_p$ 
from  small $n$ with 
$D_\perp=3.25 \times 10^{-7}~\sigma_1^2/t_0$, 
where   the jump number is of order  $10^{-4}N$  per cycle 
and is small.

\begin{figure}[t]
\includegraphics[scale=0.5]{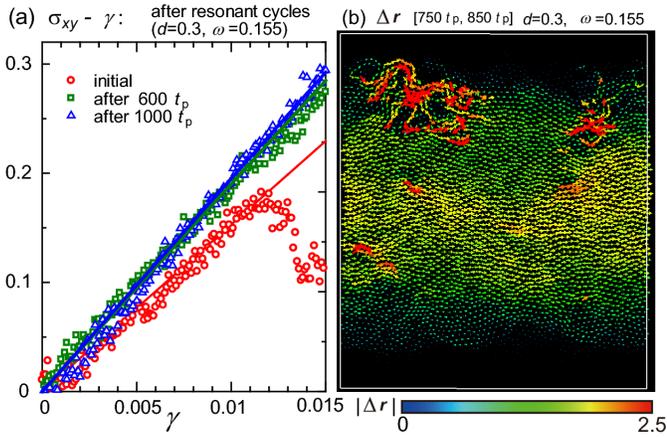}
\caption{(Color online)
(a)  Stress $\sigma_{xy}$ vs strain $\gamma$  in three states
 at $T=0.01$. Initial slope of each curve
 gives  the shear modulus $G$.
It is 16  for the initial state in 
simulation. It is 20  for  the other states
obtained by switching-off
the oscillation after  $600$ and 1000 cycles 
in   the run   of $d=0.3$ in Fig. 4.
(b)  Displacements $\Delta {\bi r}_i$
in  time interval  $[750t_p, 850t_p]$
in the run of $d=0.3$ at  $\omega=0.155$ in Fig. 4.
Here, large plastic events and system-size   sliding 
 break resonance. Particles with $|\Delta {\bi r}_i|>2$ (in red) amount to 24 
due to multiple stringlike motions. }
\end{figure}

\begin{figure}
\includegraphics[width=0.9\linewidth]{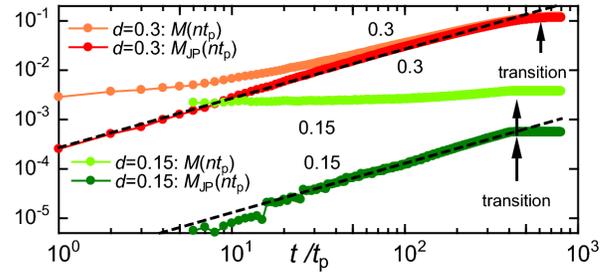}
\caption{(Color online) Stroboscopic mean square displacement 
$M(n)$ and contribution to it from the particles 
with large jumps  $M_{\rm JP}(n)$ 
as functions of $n=t/t_p <1000$ for $d=0.15$ and 0.3 
at   $\omega=0.155$.  Here, $M(n) 
\propto D_\perp n$ at large $n$  only for $d=0.3$, but 
  $M_{\rm JP}(n) \propto D_\perp n$ 
  after several cycles even for $d=0.15$. 
They cease to increase after the big 
drop (arrow). 
}
\end{figure}

{\it Summary.--} 
We have examined acoustic resonance in a 2D model glass 
under periodic  shear  with amplitude $d$ and frequency $\omega$  
applied  at a wall.  
The resonant displacements can be very  large even for 
small $d$. The damping 
arises from heterogeneous  and intermittent  plastic events.
We have found resonance  hardening 
(increase in the  shear modulus $G$), 
which could be used in technological 
applications. Here, we predict that if we increase 
$\omega$ gradually depending on $G$, we can maintain  resonance 
to achieve further hardening.

We  still  do not  understand  how the sound waves are emitted, 
deformed,   and reflected 
  in glass,  where  plastic events 
come into play at large amplitudes. See very  complex wave behaviors  
in the movie in SM \cite{Supp}. We should  further  examine  how the 
resonance saturation occurs  and 
 how  the structural changes proceed during resonance. 
We will  also report on  resonance of longitudinal sounds 
in glass by periodically changing the cell 
volume.  For   crystals and polycrystals, 
  we should investigate  how the resonance 
is influenced by  the structural defects.

\begin{acknowledgments}
This work was supported by 
funding from JSPS Kakenhi (15K05256, 
15H06263, 16H04025, 16H04034, and 16H06018).
We  would like to thank Kyohei Takae 
for valuable discussions. 
\end{acknowledgments}

\clearpage
\widetext

\setcounter{equation}{0}
\setcounter{figure}{0}
\setcounter{table}{0}
\setcounter{page}{1}

\noindent{\bf\large Supplementary Material}
\begin{center}
{\bf  \large Acoustic resonance    in  periodically sheared  glass }
\\
 
{Takeshi  Kawasaki$^1$ and Akira Onuki$^2$}\\
$^1$Department of Physics, Nagoya University, Nagoya 464-8602, Japan\\
$^2$Department of Physics, Kyoto University, Kyoto 606-8502, Japan

\end{center}

\setcounter{equation}{0}
\renewcommand{\theequation}{S.\arabic{equation}}
\renewcommand{\thefigure}{S\arabic{figure}}

\renewcommand{\bibnumfmt}[1]{[S#1]}
\renewcommand{\citenumfont}[1]{S#1}

\section{Early-stage time-evolution: caption of movie}
\vspace{-3mm}
In Fig. 1 of our Letter, we have shown resonant  growth of 
the kinetic and potential energies 
for $d=0.30$ and $\omega=0.155$. Here, in Fig. S1(a),  
we obtain a periodically deformed  state 
with small acoustic energy  
 at off-resonance at $\omega=0.1$.  
We also explain the   movie attached, 
which illustrates   time evolution  for $d=0.3$ and $\omega=0.155$ 
in the first ten cycles ($0<t<10t_p$). Depicted  are  
 the incremental changes of the particle positions, 
\be 
\Delta {\bi r}_i (t, \Delta t)= 
{\bi r}_i (t+\Delta t)-{\bi r}_i (t) \quad (t/\Delta t=0, 1, 2,\cdots, \quad 
t/t_p<10),
\en 
 where   $\Delta t=t_p/40\cong 1.0$. 
In (b), the kinetic energy deviation $\Delta K(t)$ 
consists of the oscillatory acoustic part 
and the slowly increasing thermal part due to heating, which 
are about  $ 0.01N\epsilon$ and $0.02N\epsilon$, 
respectively,  at $t/t_p\sim 10$. 
The  random thermal motions 
within narrow cages  
change on a rapid time scale of $0.1$ 
and appear as small noises in $\Delta {\bi r}_i$. 
In the movie, the  arrows  with noticeable sizes 
can be identified as  the  acoustic  velocities 
multiplied by $\Delta t$. 
\begin{figure}[h]
\begin{center}
\includegraphics[width=440pt]{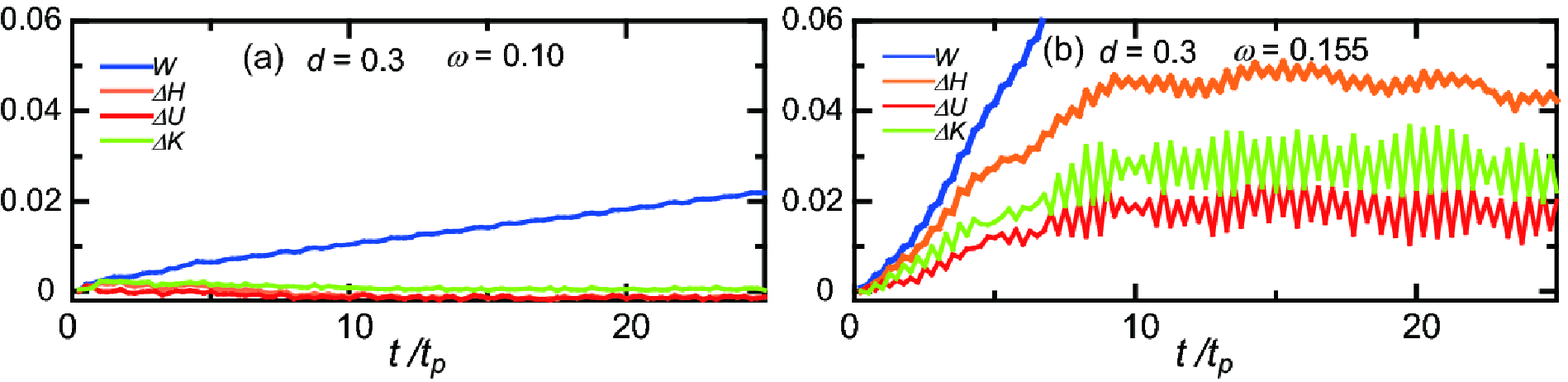}
\caption{ Deviations of 
  kinetic energy $ K$,   potential energy $U$, 
and   sum $K+U$ divided by  $N\epsilon $ with   
   $W(t)$  being   energy input
 after application of periodic shear, 
where   (a)   $\omega=0.1$ 
 with  $d= 0.30 $  at off-resonance and (b)   
$\omega=\omega_1 =0.155$ with  $d= 0.3$ at resonance. 
 In (a)  $\Delta K$ and $\Delta U$  remain small, 
while (b) corresponds to the movie attached.   
 }
\end{center}
\end{figure}

\section{Eigenmodes with rigid walls}
\vspace{-3mm}
To understand   the resonance we should also  examine   the  vibrational normal  modes from  the Hessian matrix, where  
the  eigenvectors  should vanish at $y=0$ and $L$.  To this end
we included the 500 particles connected to the  
 boundary walls (in the regions $-L/16<y<0$ and $L<y<17L/16$) 
  by the spring potentials.  We thus treated   a 2D binary mixture 
of 4500 particles in a $L \times 9L/8$ rectangle   with $L=70.2$,  
  imposing  the periodic boundary condition along the $x$ axis. 
We used the   particle configuration   obtained by cooling the  state   
  after  $ 700$ resonant cycles 
for $d=0.3$ in Fig. 4 (see the explanation of Fig. 5(a)). 
This linear  analysis itself is of interest, because  
the periodic boundary condition has been imposed  along  all the axes in 
the previous calculations of the vibrational modes (see Refs.$[22-26]$ 
in our Letter).

\begin{figure*}[t]
\includegraphics[scale=0.9]{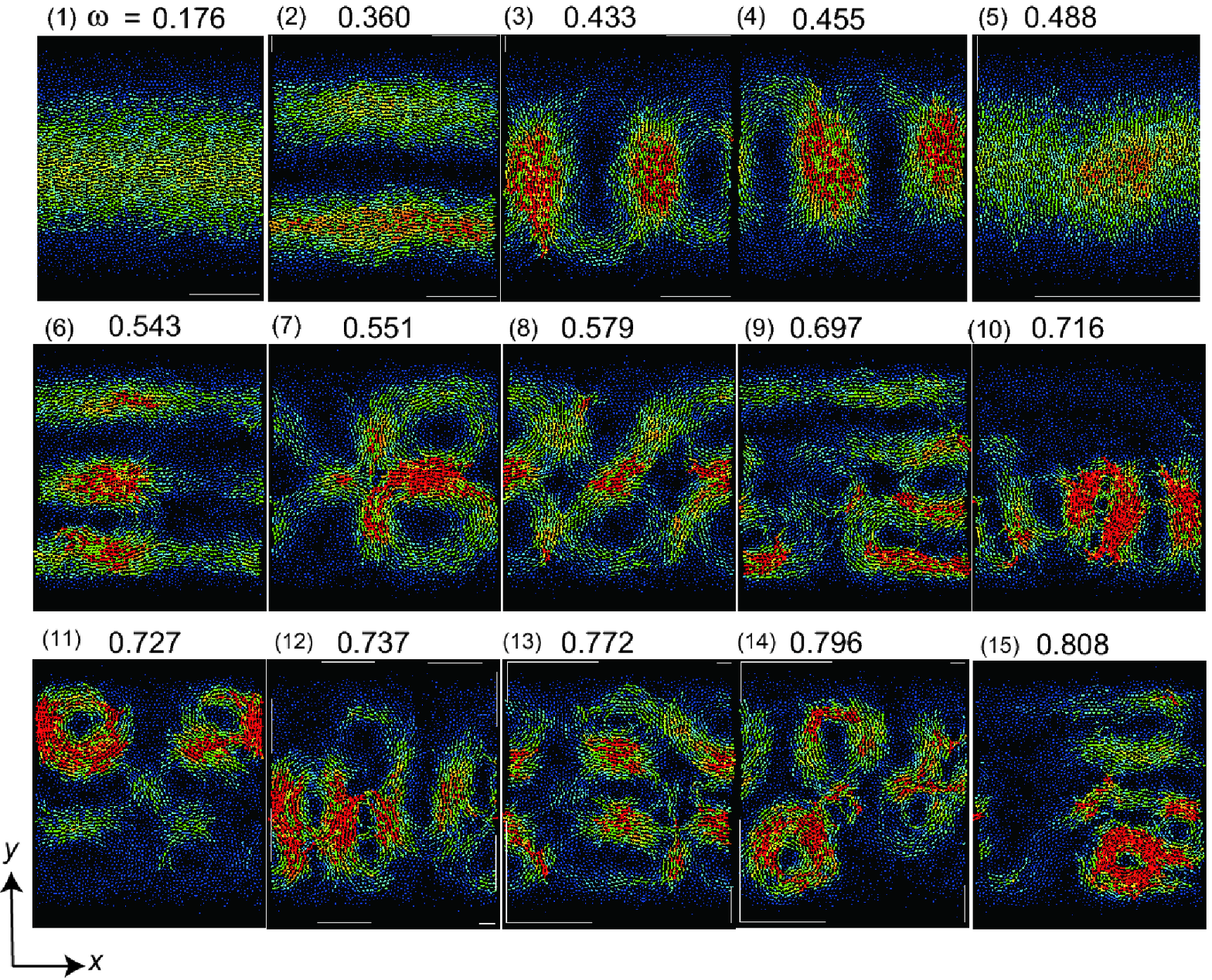}
\caption{ First 15  normal  modes 
obtained from  the Hessian matrix with the rigid 
boundary condition at $-L/16<y<0$ and $L<y<17L/16$ 
and the periodic boundary condition along the $x$ axis. 
Colors of the particles  represent the displacement 
magnitude  $|{\bi u}_i|$ of the corresponding eigenmode.  
Particle positions in the Hessian matrix  are obtained from cooling 
the state  at $t=700 t_p$ with $d=0.3$ in  Fig. 4(a). The first, second, 
and sixth shear modes are amplified in Fig. 3 in our Letter. }
\end{figure*} 

Our system is sufficiently large  such that 
the first few extended sound modes 
have lower  frequencies $\omega$ than  those of 
the quasi-localized vibrations \cite{SchS}. 
 Thus, at low $\omega$, 
we can compare the eigenmodes from the Hessian matrix and  the elastic 
modes    from the  linear elasticity (EMs). 
In the latter, the 2D displacement  ${\bi u}= (u_x, u_y)$   obeys \cite{LandauS}  
\be 
-\rho \omega^2 {\bi u}=B \nabla(\nabla\cdot{\bi u}) + G \nabla^2{\bi u},
\en  
where $\rho$ is the mass density,  
$B$ is the bulk modulus, and $G$ is the shear modulus. 
Dissipation is neglected here. We solve 
this equation assuming the sinusoidal $x$ dependence 
${\bi u} \propto e^{ik_x x}$ and the boundary condition  ${\bi u}={\bi 0}$ 
at $y=0$ and $L$ in  the complex number representation. 

First, for  $k_x=0$, we obtain  the transverse and   longitudinal EMs:  
\be 
{\bi u}_\ell^{\rm T}
=\sin (\pi \ell y/L) {\bi e}_x,\quad  
{\bi u}_\ell^{\rm L}=\sin (\pi \ell y/L) {\bi e}_y 
\quad (\ell=1, 2, \cdots),
\en  
where ${\bi e}_x$ and ${\bi e}_y$ are   the unit vectors  
along the $x$ and  $y$ axes, respectively. In terms of 
the sound speeds 
$c_\perp= (G/\rho)^{1/2}$ and   $c_\parallel= [(B+G)/\rho]^{1/2}$, 
their eigenfrequencies  are given    by 
\be 
\omega_\ell= \pi\ell c_\perp/L, 
\quad \omega_\ell' = \pi\ell c_\parallel/L \quad (\ell=1, 2, \cdots).  
\en 
Here,  $\pi$ is  replaced by $2\pi$ in 
 the periodic boundary condition  along the $y$ axis. 
Second, let   $k_x= 2\pi/L$ in Eq. (S2). 
In this case, $\bi u$ represents a mixed EM, 
since it can be expressed as 
\be 
u_x= -\nabla_y H -ik_x J ,\quad 
u_y=ik_x H-\nabla_y  J, 
\en 
where $\nabla_y=\p/\p y$. The  two functions $H(y)$ and 
$J(y)$  satisfy 
 $\nabla_y^2 H= (k_x^2- \omega^2/c_\perp^2)H$ and 
 $\nabla_y^2 J= (k_x^2- \omega^2/c_\parallel^2)J$. 
This decomposition of $\bi u$ 
into the transverse and longitudinal parts 
can be used to  calculate  the Rayleigh surface wave \cite{LandauS}. 
In the range  $\omega_2 <\omega<\omega_2' $ we set  
\be 
H=H_0 \cos[q_\perp( y-L/2)],\quad 
J= J_0 \sinh [q_\parallel( y-L/2)]
\en  
where  $(q_\perp/k_x)^2= (\omega/\omega_2)^2-1$ and 
$(q_\parallel/k_x)^2= 1- (\omega/\omega_2')^2$ with 
$H_0$ and $J_0$ being constants. 
Here, $\omega$ is determined by    
$q_\perp q_\parallel/k_x^2 
 = -  \cot(q_\perp L/2)\tanh(q_\parallel L/2)$, 
which gives  $\omega/\omega_2=1.17$ 
for   $c_\parallel/c_\perp=3$, for example. 

In Fig. S2, we display the first 15 eigenmodes of the Hessian matrix, 
which vanish at $y=0$ and $L$. 
 The first two modes correspond 
to  those of  transverse EMs $(\ell=1,2)$. 
However, their   frequencies $0.176$ and $0.360$ are  
somewhat higher than the resonant ones $0.155$  and $0.323$ 
 in Figs. 2 and 3. The third and fourth  modes are roughly proportional to 
$\cos(2\pi x/L +\alpha)$ and $\sin (2\pi x/L +\alpha)$ 
with $\alpha$ being a constant, so they   correspond to 
the mixed EM ($\propto e^{2\pi i x/L}$)  in Eq. (S5) with 
$\omega/\omega_2\sim 1.2$. 
In fact, $u_x$ is odd and $u_y$ is even 
as functions of $y- L/2$ for these modes. 
In the fifth mode, 
all the displacements of the particles are upward, so it 
 corresponds to  the first longitudinal EM, 
leading to  $c_\parallel\cong 11$ from  $\omega= \pi c_\parallel/L$. 
The sixth and ninth  ones correspond to the 
third and fourth transverse  EMs  $(\ell=3$ and 4), but they  largely 
vary in   the $x$ direction.   
The seventh and eighth ones look similar to the sixth one, 
but are amplified at the middle ($y\sim L/2$). 
From the sixth to ninth modes, 
 the excited regions form  long stripes. 
The other ones  vary in space both 
in the $x$ and $y$ directions in quasi-localized 
 manners, where  mesoscopic regions of large 
displacements are distinctly separated but 
are weakly connected \cite{SchS}. 
In our case, all the eigenvectors from the Hessian 
matrix  are extended in  the whole cell. 
  This aspect has not been well  studied.

We also calculated  the eigenvectors 
for the particle positions in  the initial state of simulation  and  
for those after 1000 cycles for $d=0.3$ in Fig. 4. 
The first 6  eigenvectors in these cases 
are nearly the same as those in Fig. S2, but  the higher quasi-localized modes 
are significantly  different. We can see that  
the quasi-localized modes  
  sensitively depend on the details of the  particle 
configurations. In the resonant states, 
 they  are also excited and mixed with the 
primary shear wave mode, 
but they  vary in time with the structural changes.

\section{Periodic solution from linear elasticity }
\vspace{-3mm}

Let us calculate the periodic shear displacement 
 $u_x(y,t)$ from  the linear elasticity  
 without damping \cite{LandauS}, which can be realized 
after many cycles at off-resonance. 
Imposing the boundary conditions 
$u_x(0,t)=0$ and $u_x(L,t)= -d \sin (\omega t)$ in Eq. (2), 
we obtain 
\be 
u_x=-d \sin(\omega t)\sin (\omega y/c_\perp)/\sin(\omega L/c_\perp),  
\en 
which  diverges as $\omega \to \omega_\ell= \pi \ell c_\perp/L$, 
so we assume    $\omega\neq \omega_\ell$.  If    $\omega\ll \omega_1$, 
this continuum  expreesion is consistent with 
 our simulation  data far from the boundary walls on the average. 
Here, the acoustic energy  $E_a$  is the space integral of  
$ \rho ({\partial  u_x}/{\partial t})^2/2+ G 
(\partial u_x/\partial y)^2/2$ in the linear elasticity.  Therefore,  if 
$E_a$ is   averaged over one cycle, we obtain    
\be 
\av{E_a} = \frac{1}{t_p} 
\int_0^{t_p} dt E_a(t)= \rho L^2 [d\omega/2\sin (\omega L/c_\perp )]^2   \quad 
(\omega\neq \omega_\ell).
\en 
For  our model system, this  gives 
$\av{E_a} \cong 2.1 \times 10^{-5}N\epsilon  
/(\omega/\omega_\ell-1)^{2}$ close to resonance 
for  $d=0.15$, where the coefficient is very small.
However, as Fig. 2 of our Letter indicates, 
the above expression holds only for $\omega\ll \omega_1$. 
Here, as $\omega \to 0$,  $\av{E_a}$ 
tends to $ G d^2/4$, where $G=\rho c_\perp^2$ 
is the   shear modulus in the limit of low  frequency 
and long wavelength.

\section{Intermittency and collective motions}
\vspace{-3mm}

\begin{figure}
\begin{center}
\includegraphics[width=440pt]{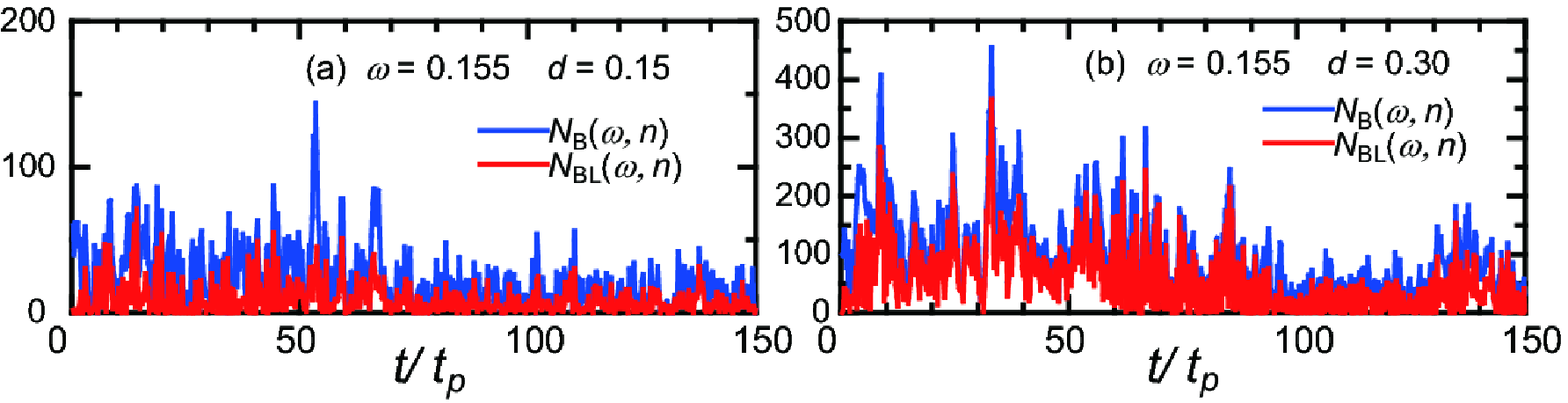}
\caption{ 
 Particle  numbers $N_{\rm B}(\omega, n)$ 
with broken bonds (in blue)  and    $N_{\rm BL} (\omega, n)$ 
with large jumps (in red) in time interval $[nt_p,(n+1)t_p]$ 
vs  $n=t/t_p$ at  $\omega=0.155$ in the range $n=t/t_p<150$,  where 
$d$ is    (a) $0.15$ and (b) 0.3. 
 }
\end{center}
\end{figure}

In Fig. 4(a), we have plotted the particle  number 
$N_{\rm B}(\omega, n)$ with broken bonds in the $n$-th cycle   
 at the first resonance   $\omega=0.155$ for   $d=0.3$ 
in the time range $n= t/t_p<1300$. In Fig. S3, we  display  
its intermittent time evolution in the shorter range $n= t/t_p<150$  
for $d=0.15$ and 0.3. Here, we should note 
 that  a considerable  fraction of these  particles with 
broken bonds   return to their original positions in subsequent  cycles, 
 as has been reported  in the off-resonant 
 situations (see Refs.$[10-17]$ in   our Letter).  
Hence,   we also  consider  the particles    with large  
displacement with  $|{\bi r}_j(n t_p+t_p)-{\bi r}_j(nt_p) | >0.8$, 
as in Fig. 6 in  our Letter  \cite{KawasakiPRES}. 
These  particles   have irreversibly  escaped from   cages  
and have also  broken bonds in  the n-th cycle, so 
 their number   ${N}_{\rm BL}(\omega, n)$ is   
smaller than or equal to   ${N}_{\rm B}(\omega, n)$ in Fig. S3.
Note that their jump motions give rise to 
translational   diffusion  as in Fig. 6.

In Fig. S4, we show snapshots of 
the irreversible displacements 
$\Delta r_i = 
{\bi r}_i (nt_p+t_p)-{\bi r}_i (nt_p)$ 
for $d=0.3$ and $\omega=0.155$ in 
 four consecutive cycles, where  $n= t/t_p$ is 
(a) 23, (b) 24, (c) 25 and  (d) 26. 
The distributions of the particles with broken bonds 
(in white)  demonstrate intermittent  fluctuations of 
the plastic events in successive cycles. 
Remarkably, in (a), (b), and (d), 
we can  see large-scale collective   motions   
with considerably large displacements ($\sim  0.5$ in (a)), 
while  in (c) such collective motions are inconspicuous. 
During  these cycles, the system remains at resonance. In  Fig. 5(b), 
 we have shown  system-size sliding along the $x$ axis 
at $n\sim 800$ in the same  run, 
which breaks  resonance.  Thus, large-scale collective motions 
of various sizes  appear  intermittently  together with   
plastic events. 
We note that they may be treated  as elastic deformations  
away from the particles with broken bonds. Indeed, long-range elastic strains 
 have been calculated  around local plastic events in  glass 
\cite{AnaelS}, which are similar to the Eshelby  strains 
 around precipitates in   metallic  alloys.

\begin{figure}
\includegraphics[width=1\linewidth]{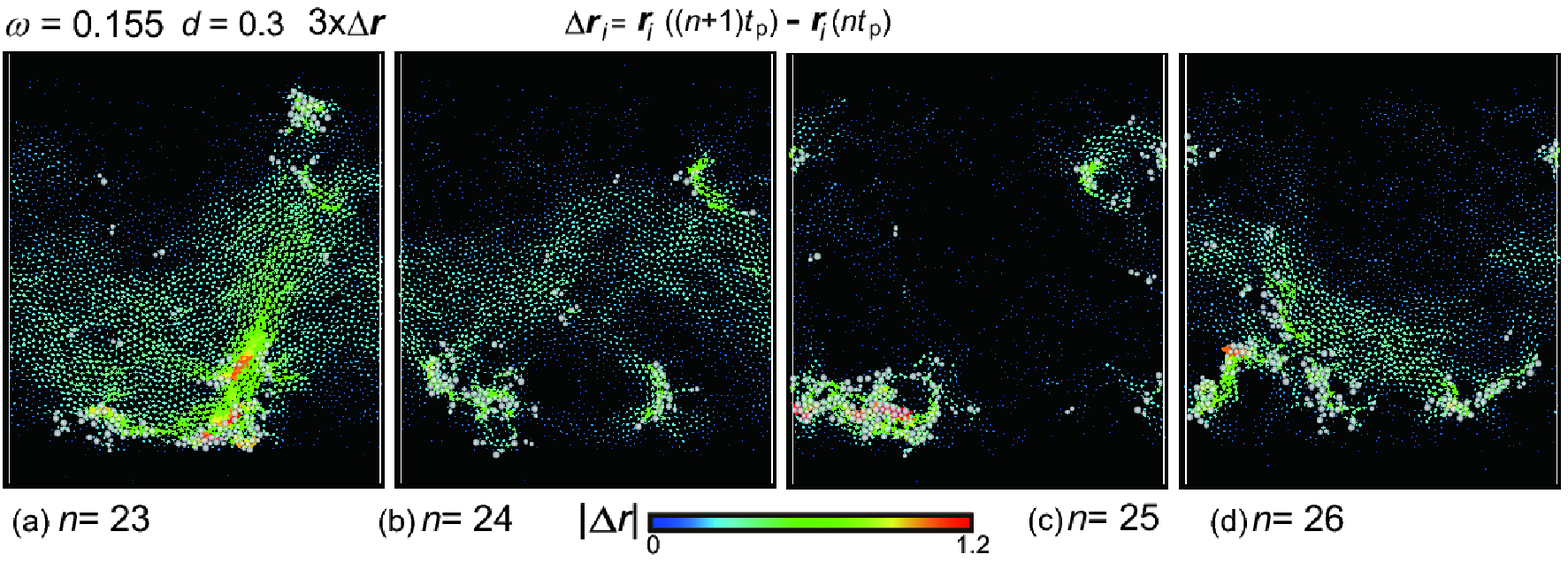}
\caption{ Snapshots of irreversible 
displacements $\Delta {\bi r}_i={\bi r}_i(nt_p+t_p)-{\bi r}_i(nt_p)$ 
for consecutive four cycles  with $n= t/t_p=23, 24, 25$ and $26$, 
where   $\omega=0.155$ and $d=0.3$. 
They  largely  fluctuate, because  collective 
  plastic events occur intermittently. 
Particles with broken bonds are drawn in while circles. 
We can see repeated plastic events in the lower region. 
Large-scale collective motions are also shown. 
 }
\end{figure}

\section{Excitation of longitudinal waves}
\vspace{-3mm}

\begin{figure}[t]
\includegraphics[width=0.82\linewidth]{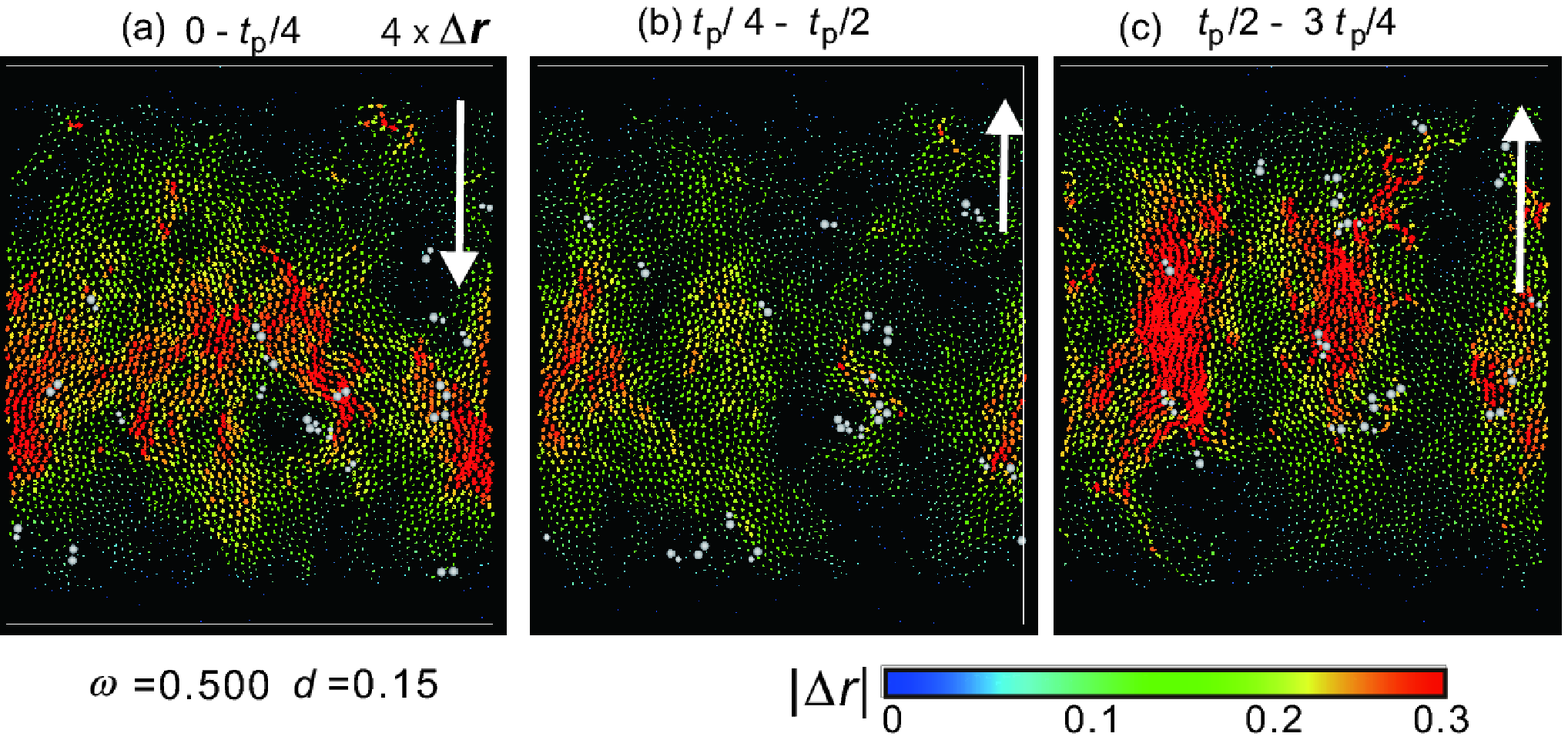}
\caption{ Resonance of  longitudinal sounds  
 induced by small periodic  shear, 
where $\omega$ is increased in a stepwise 
manner up to $0.5$ 
with   $d=0.15$. Depicted are   
$\Delta{\bi r}_j = {\bi r}_j(t_1+t_p/4)
-{\bi r}_j(t_1)$ (multiplied by 4) in consecutive 
quarter cycles  with width $t_p/4$, 
where $t_1$ is (a) $225t_p$, (b)$(225+1/4)t_p$, 
and (c) $(225+1/2)t_p$. 
Averages of $\Delta {\bi r}_i=(\Delta x_i, \Delta y_i) $ 
over all the particles are  
(a) $(0.07, -0.14)$, (b) $(0.003, 0.095)$, 
and (c) $(-0.02, 0.14)$ (white arrows). 
  }
\end{figure}

In  our  Letter, we  
started  with the same initial state and applied 
the periodic shear in Eq. (2) fixing  $d$ and 
$\omega$ in each simulation run. We also performed simulation 
by increasing  $\omega$ slowly in a stepwise manner at each   
fixed $d$, where  we found considerably 
different resonance behavior at relatively high $\omega\gs \omega_2$. 
In particular, we realized resonance  of longitudinal sounds 
at $\omega \cong \pi  c_\parallel/L\sim 0.5$ 
with $c_\parallel \sim 10$. 
In fact, in  Fig. S5,   we show 
 amplified compression and expansion along the $y$ axis 
vanishing  at the walls for $\omega=0.5$  with $d=0.15$. 
 We can see that    the particle displacements 
 are mostly downward in (a)  and  upward in (b) and (c), though  
they are partially  transverse 
varying along the $x$ axis. 
Therefore,  the resonant behaviors  
are so  complex  in glass  such that  they even depend 
  on the simulation path (protocol). In future work we 
will apply  periodic dilation to glass to induce 
longitudinal wave  resonance.


\begin{thebibliography}{99}
\bibitem{input} 
The  energy input $W(t)$  is 
defined  by $d W(t)/dt 
=  \sum_{j\in {\rm top} } {\bi v}_j(t)\cdot{\bi f}_j(t)$, 
where    particle  $j$ is  in the upper layer,  
  ${\bi v}_j$ is its velocity, and $ {\bi f}_j$ is the force on  $j$.
\bibitem{comment1} If a steady shear with rate $\dot\gamma$ 
is applied in  glass, the bond breakage number 
is about $ {\dot\gamma} \Delta t N$ in a time interval with 
width $\Delta t$ on the average \cite{Yamamoto}. 
Then,  the energy dissipation 
 is of order $\epsilon {\dot \gamma} N\Delta t$ due 
to plastic events.  Equating this with 
 $\eta(\dot{\gamma}){\dot\gamma}^2V\Delta t$, 
we find  the nonlinear viscosity  
$\eta(\dot{\gamma})\sim \epsilon n /{\dot\gamma}$, where 
$n=N/V$ is the density. In our  case, 
we make similar arguments in explaining   Fig. 3.
\bibitem{comment2} If $T$  was raised homogeneously 
after the big drop 
in the run of $d=0.3$ in Fig. 4, 
 $G$ was decreased to $17$ at $ T=0.05$ and 
to 16 at $ T=0.1$, where $G= 20$ at 
$T=0.01$.
\bibitem{diffusion} The  diffusion constant  
along the $x$ axis $D_\parallel$ is  a few times larger 
than $D_\perp$ at $\omega=\omega_1$ (not shown here).
\end{thebibliography}
\end{document}